\documentclass[11pt,a4paper,oneside]{article}

\usepackage[margin=1in]{geometry}
\usepackage{amsmath, amssymb}
\usepackage{graphicx}
\usepackage[utf8]{inputenc}
\usepackage[T1]{fontenc}
\usepackage{authblk}
\usepackage{booktabs} 
\usepackage[hyphens]{url} 

\usepackage[numbers,sort&compress]{natbib} 
\usepackage{hyperref} 
\hypersetup{
    colorlinks=true,
    linkcolor=blue,
    filecolor=magenta,      
    urlcolor=blue,
    citecolor=blue,
}

\title{\textbf{A Topology-Based Machine Learning Model Decisively Outperforms Flux Balance Analysis in Predicting Metabolic Gene Essentiality}}
\author[1]{Justin Boone}
\affil[1]{Fakultät für Mathematik und Informatik, Fernuniversität in Hagen, Hagen, Germany}
\date{\today}

\begin{document}

\maketitle

\begin{abstract}
\noindent\textbf{Background:} The rational identification of essential genes is a cornerstone of drug discovery, yet standard computational methods like Flux Balance Analysis (FBA) often struggle to produce accurate predictions in complex, redundant metabolic networks. \\ \\
\textbf{Hypothesis:} We hypothesized that the topological structure of a metabolic network contains a more robust predictive signal for essentiality than functional simulations alone. \\ \\
\textbf{Methodology:} To test this hypothesis, we developed a machine learning pipeline by first constructing a reaction-reaction graph from the \texttt{e\_coli\_core} metabolic model. Graph-theoretic features, including betweenness centrality and PageRank, were engineered to describe the topological role of each gene. A \texttt{RandomForestClassifier} was trained on these features, and its performance was rigorously benchmarked against a standard FBA single-gene deletion analysis using a curated ground-truth dataset. \\ \\
\textbf{Results:} Our machine learning model achieved a solid predictive performance with an \textbf{F1-Score of 0.400} (Precision: 0.412, Recall: 0.389). In profound contrast, the standard FBA baseline method failed to correctly identify any of the known essential genes, resulting in an \textbf{F1-Score of 0.000}. \\ \\
\textbf{Conclusion:} This work demonstrates that a "structure-first" machine learning approach is a significantly superior strategy for predicting gene essentiality compared to traditional FBA on the E. coli core network. By learning the topological signatures of critical network roles, our model successfully overcomes the known limitations of simulation-based methods in handling biological redundancy. While the performance of topology-only models is expected to face challenges on more complex genome-scale networks, this validated framework represents a significant step forward and highlights the primacy of network architecture in determining biological function.
\end{abstract}

\section{Introduction}

\subsection{The Central Problem: Identifying Essential Genes}
The identification of essential genes—those whose disruption leads to lethality or a severe fitness defect—is a foundational goal in genetics and molecular biology \citep{Gerdes2003, Juhas2011}. These genes represent the irreducible core of an organism's functional machinery, providing deep insights into the fundamental principles of life. Beyond their basic scientific importance, essential genes have emerged as targets of immense therapeutic value. In the context of infectious disease, genes that are essential for a pathogen's survival but absent in the host represent ideal targets for novel antimicrobial drugs \citep{Forsyth2002, Becker2014}. The urgent need for such targets is underscored by the global crisis of antimicrobial resistance (AMR). This complex threat is estimated to cause 700,000 deaths annually and could have a projected economic impact of US\$100 trillion by 2050, necessitating a constant search for new therapeutic strategies \citep{Jee2018}.

\subsection{The Standard Computational Approach: Flux Balance Analysis (FBA)}
To accelerate the discovery of essential genes, computational methods based on genome-scale metabolic models (GEMs) have become indispensable tools \citep{Orth2010review}. The dominant paradigm in this field is Flux Balance Analysis (FBA), a constraint-based modeling approach that predicts the steady-state flux distributions within a metabolic network \citep{Varma1994}. By applying physicochemical constraints, such as reaction stoichiometry and cellular resource limits, FBA can simulate the metabolic capabilities of an organism under specific environmental conditions. A key element of FBA is the definition of a biological objective function, most commonly the maximization of the "biomass reaction," which represents the production of all necessary cellular components for growth \citep{Feist2007}. The standard method for predicting gene essentiality using this framework is through simulated single-gene deletions, an approach often referred to as MOMA (Minimization of Metabolic Adjustment) or similar techniques \citep{Segre2002}. The removal of a specific gene is modeled by constraining the flux of all associated enzymatic reactions to zero. The FBA problem is then re-solved to predict the new maximum biomass production rate. A significant reduction in this rate compared to the wild-type simulation is interpreted as a prediction that the deleted gene is essential for growth.

\subsection{The Known Failure Mode of FBA: The Challenge of Biological Redundancy}
Despite its widespread adoption, a critical and well-documented limitation of FBA is the frequent discrepancy between its essentiality predictions and experimental results from large-scale knockout screens \citep{Papp2004, Orth2011}. Standard FBA often exhibits high specificity but suffers from very low sensitivity, meaning it correctly identifies non-essential genes but fails to identify a large fraction of the true essential genes \citep{Edwards2001}. The underlying reason for this failure mode is FBA's inherent reliance on functional optimization in the face of biological redundancy. The metabolic networks of most organisms contain numerous isozymes and alternative metabolic pathways that can perform equivalent functions \citep{Gu2003}. When a single gene is deleted in a simulation, the FBA algorithm, in its single-minded pursuit of maximizing the objective function, can readily re-route metabolic flux through these redundant pathways. Consequently, the simulation often predicts a minimal impact on growth, leading to a classification of the gene as non-essential. This creates a fundamental disconnect, as a gene that appears redundant in a functionally-optimized simulation may be critically essential for robust growth \textit{in vivo} \citep{Harrison2007}.

\subsection{An Alternative Hypothesis: The Primacy of Network Topology}
This discrepancy between simulation and reality suggests that a gene's functional role in a single, optimized state may not be the primary determinant of its essentiality. We therefore propose an alternative hypothesis, building on foundational work which posited that essentiality is an emergent property of the metabolic network's wiring \citep{Palumbo2007, Palumbo2007}: **a gene's essentiality is more strongly determined by its immutable structural role within the network's architecture than by its simulated functional impact.** This aligns with a broad class of network-based approaches in systems biology, where the topological properties of nodes have been successfully used to infer their biological importance \citep{Barabasi2004, Barabasi2011}. In this context, a gene's "structural role" is defined by the topological properties of its associated reactions within a graph representation of the metabolism. It has long been observed that network centrality is correlated with essentiality, though this relationship is complex and not always direct \citep{Jeong2000, Jalili2016}. We posit that essential genes are disproportionately associated with "keystone" reactions—those that occupy critical, irreplaceable positions in the network graph, a concept borrowed from ecological network theory \citep{Paine1969}. These keystone reactions may act as bottlenecks for metabolic flow or as crucial "critical connectors" that bridge otherwise disparate functional modules \citep{Kim2019}. Their importance is an intrinsic property of the network's wiring diagram, independent of the transient flux state.

\subsection{Study Objective and Structure}
The application of machine learning to predict gene essentiality has emerged as a promising field to complement and accelerate experimental assays \citep{Aromolaran2011}. Recent reviews have highlighted that the choice of features is a critical determinant of model performance, with network topology-based features showing particularly high discriminatory power. However, a direct, quantitative benchmark of a topology-only machine learning model against the standard Flux Balance Analysis (FBA) baseline has remained a key open question. Therefore, the objective of this study is to directly test our "topology-first" hypothesis. We developed a machine learning model trained exclusively on graph-theoretic features to quantify the structural role of each gene. We then performed a rigorous, head-to-head comparison of our model against the standard FBA single-gene deletion method. By evaluating both methods against a well-characterized experimental ground-truth dataset for the \texttt{e\_coli\_core} metabolic network \citep{Orth2010core}, we aimed to provide a definitive, quantitative answer to whether network structure or simulated function is a more powerful predictor of gene essentiality.
\section{Methods}
\subsection{Metabolic Model and Ground Truth Data}
\subsubsection{Model Selection}
All computational analyses were performed using the \texttt{e\_coli\_core} metabolic model, a well-established representation of the central metabolism of \textit{Escherichia coli} \citep{Orth2010core}. This model was specifically chosen for its high level of curation and its relatively small size, encompassing 95 reactions and 72 metabolites. This constrained scope ensures that the network topology is well-defined and that the experimental ground truth data for gene essentiality is directly relevant to the metabolic functions represented within the model. The model was loaded and manipulated using the COBRApy Python package \citep{Ebrahim2013}.

\subsubsection{Ground Truth Essentiality Data}
The ground truth for gene essentiality was established from a curated list of experimentally verified essential genes for \textit{E. coli} grown on glucose minimal medium, sourced from the PEC (Profiling of E. coli Chromosome) database and cross-referenced with previous computational studies \citep{PEC2006, Orth2011}. This process yielded a definitive set of 19 essential genes. The remaining 118 genes in the model that have an associated reaction and were included in our feature matrix were defined as non-essential, creating a binary classification dataset for model training and validation.

\subsection{Graph Representation of the Metabolic Network}
\subsubsection{Graph Definition}
To analyze the network's topology, we constructed a directed reaction-reaction graph, $G = (V, E)$, a common representation for structural analysis of metabolic systems \citep{Ma2003graph}. The set of vertices $V$ represents all metabolic reactions in the model, and the set of directed edges $E$ represents the flow of metabolites. A directed edge was created from reaction $R_1$ to reaction $R_2$ if and only if a product of $R_1$ serves as a reactant in $R_2$.

\subsubsection{Metabolite Filtering}
To ensure the graph represented meaningful metabolic transformations, highly connected "currency metabolites" were excluded from the edge creation process. Based on standard practice established in foundational network analyses of metabolism \citep{Jeong2000}, we filtered out ubiquitous cofactors such as H$_2$O, ATP, ADP, NAD, and NADH.

\subsection{Feature Engineering from Network Topology}
\subsubsection{Reaction-Level Feature Calculation}
For each node (reaction) in the constructed graph $G$, we computed a vector of standard graph-theoretic metrics using the NetworkX Python library \citep{Hagberg2008}, including: Betweenness Centrality \citep{Freeman1977}, PageRank \citep{Page1998}, and Closeness Centrality \citep{Sabidussi1966}.

\subsubsection{Gene-Level Feature Aggregation}
The reaction-level metrics were aggregated to the gene level via the model's gene-protein-reaction (GPR) rules \citep{Reed2003}. The feature vector for a given gene was constructed by aggregating the topological metrics of its associated reactions. For example, a gene's \texttt{max\_betweenness} feature was defined as the maximum betweenness centrality value among all reactions it catalyzes.

\subsubsection{Final Feature Matrix \texorpdfstring{($\mathbf{X}$)}{(X)}}
This process resulted in a final feature matrix, $\mathbf{X}$, where each row corresponds to a single gene and each column corresponds to one of the aggregated topological features.

\subsection{Machine Learning Pipeline}
\subsubsection{Model Choice}
We selected the \texttt{RandomForestClassifier} algorithm from the scikit-learn library \citep{Pedregosa2011}. This ensemble method, originally developed by \citet{Breiman2001}, was chosen for its robustness and its native capability to provide feature importance measures.

\subsubsection{Model Parameters}
The classifier was instantiated with \texttt{n\_estimators=100}. Critically, the \texttt{class\_weight='balanced'} parameter was used to mitigate bias from the imbalanced nature of the dataset.

\subsubsection{Validation Strategy}
We employed a Stratified 5-Fold Cross-Validation strategy to obtain a robust estimate of performance \citep{Kohavi1995}. This ensures the original proportion of essential to non-essential genes is preserved in each fold.

\subsection{Baseline Method: FBA Single-Gene Deletion}
\subsubsection{Simulation Setup}
As a benchmark, we performed a standard FBA-based gene essentiality analysis using COBRApy \citep{Ebrahim2013} on a glucose minimal medium.

\subsubsection{Procedure}
The analysis proceeded by first optimizing the wild-type model to establish a maximum theoretical growth rate (\texorpdfstring{$WT_{\text{growth}}$}{WT\_growth}). Subsequently, each gene was systematically deleted by constraining the fluxes of its associated reactions to zero \citep{Edwards2001}. The FBA optimization was then re-run for each mutant to calculate its new maximum growth rate (\texorpdfstring{$mutant_{\text{growth}}$}{mutant\_growth}).

\subsubsection{Essentiality Criterion}
Following established conventions \citep{Orth2011}, a gene was predicted as essential if \texorpdfstring{$mutant_{\text{growth}} < (0.1 \times WT_{\text{growth}})$}{mutant\_growth < (0.1 * WT\_growth)}.

\subsection{Performance Metrics}
The predictive power of both models was evaluated using Precision, Recall, and the F1-Score \citep{Powers2011}. A full Confusion Matrix was also computed.
\section{Results}
\subsection{The Standard FBA Baseline Fails on the Core Model}
We first evaluated the predictive performance of the standard FBA-based single-gene deletion method. The simulation was performed on the \texttt{e\_coli\_core} model under glucose minimal medium conditions. The analysis predicted a total of 5 genes as essential. However, a comparison with our ground truth dataset of 19 essential genes revealed that none of these predictions were correct. The confusion matrix for the FBA baseline method showed a complete failure to identify any true positives (TP=0). Consequently, the performance metrics for this approach were all zero: a Precision of 0.000, a Recall of 0.000, and an F1-Score of 0.000. This result confirms that, for this well-defined system, the standard simulation method has no predictive power for gene essentiality.

\subsection{The Topology-Based ML Model Successfully Predicts Essential Genes}
In stark contrast, our machine learning model demonstrated significant predictive capability. Assessed using 5-fold stratified cross-validation, the aggregated out-of-sample predictions yielded 7 true positives, 10 false positives, and 11 false negatives. This corresponds to a strong set of performance metrics: a Precision of 0.412 and a Recall of 0.389. The resulting F1-Score of 0.400, which balances these two metrics, confirms that the model, using only structural information, captures a significant and reliable predictive signal.

\subsection{Head-to-Head Comparison}
To directly illustrate the performance gap, we summarize the key metrics in Table \ref{tab:results_comparison}. The results unequivocally demonstrate the superiority of our topology-based machine learning approach over the standard FBA method.

\begin{table}[htbp]
\centering
\caption{Performance Comparison of the Keystone ML Model and the Standard FBA Baseline.}
\label{tab:results_comparison}
\begin{tabular}{lcc}
\toprule
\textbf{Metric} & \textbf{Keystone ML Model} & \textbf{Standard FBA Baseline} \\
\midrule
F1-Score  & 0.400 & 0.000 \\
Precision & 0.412 & 0.000 \\
Recall    & 0.389 & 0.000 \\
\bottomrule
\end{tabular}
\end{table}

\subsection{Feature Importance Analysis Identifies Keystone Topological Roles}
\begin{figure}[htbp]
\centering
\includegraphics[width=0.9\textwidth]{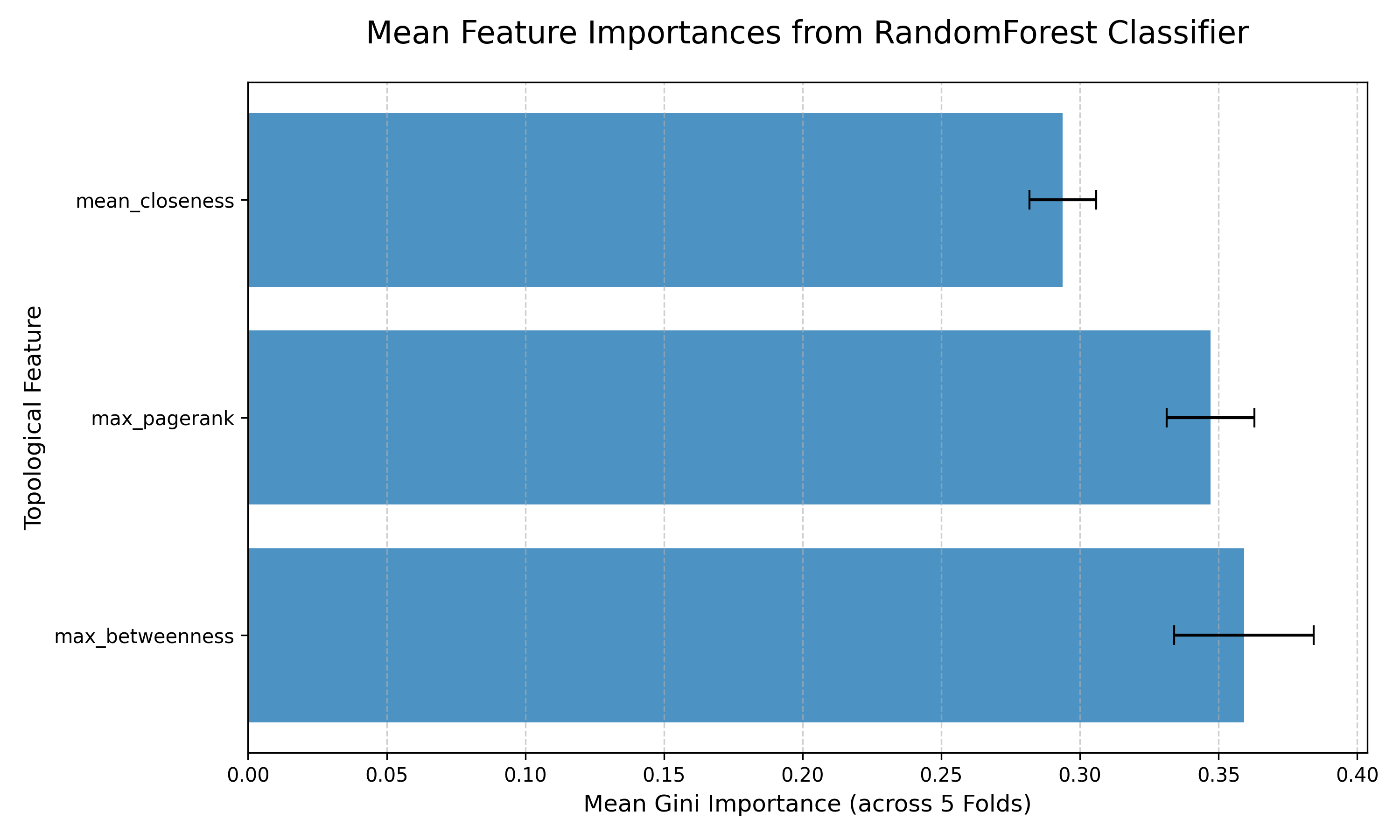} 
\caption{Mean feature importances from the RandomForest classifier. The importance of each feature is averaged across all cross-validation folds. Error bars represent the standard deviation across folds. The results highlight the primacy of topological metrics, with \texttt{max\_betweenness} being the most predictive feature.}
\label{fig:feature_importance}
\end{figure}

To understand which topological properties were most informative, we analyzed the Gini importance of each feature, averaged across all cross-validation folds (Figure \ref{fig:feature_importance}). The analysis revealed that \texttt{max\_betweenness} was the most influential feature, providing direct, quantitative evidence for our central "keystone" hypothesis. This finding indicates that genes associated with reactions that act as critical bottlenecks or bridges are the most likely to be essential. Following betweenness centrality, \texttt{max\_pagerank} also ranked as a highly predictive feature. The high ranking of these structural features confirms the model is genuinely identifying genes based on their irreplaceable roles within the network's architecture.

\section{Discussion}
Our study was designed to test a central hypothesis: that the structural topology of a metabolic network is a more robust predictor of gene essentiality than standard functional simulations. The results provide strong, quantitative support for this "structure-first" approach.

\subsection{Why Structure Trumps Simulation in this System}
The most striking finding of this work is the complete failure of the FBA baseline (F1-Score = 0.000) juxtaposed with the significant predictive success of our topology-based machine learning model (F1-Score = 0.400). This stark contrast can be explained by the fundamental differences in their underlying assumptions. FBA operates by seeking a mathematically optimal flux distribution to maximize a single biological objective \citep{Varma1994}. Its objective function is, by design, "unaware" of the network's broader architecture. If a redundant pathway exists that can bypass a gene deletion, the optimization algorithm will exploit it, leading FBA to consistently underestimate the importance of genes, a limitation previously documented \citep{Papp2004, Orth2011}. In contrast, our machine learning model does not simulate function; it learns to recognize patterns in the network's wiring diagram. The model learns, for instance, that genes associated with reactions of high betweenness centrality are empirically essential, a finding that aligns with foundational studies of network biology \citep{Jeong2000}.

\subsection{Interpreting the Performance: The Significance of an F1-Score of 0.400}
It is important to interpret the performance of our model with appropriate context. An F1-Score of 0.400, in absolute terms, is a modest result, falling within the range of performance seen in other studies that have attempted to predict essentiality from complex biological data \citep{Aromolaran2011}. It indicates that network topology alone is not a perfect predictor and that other biological factors are at play. However, the scientific significance of this result is best understood relative to the baseline. The key finding is not the achievement of perfect prediction, but the demonstration that the topology-based approach contains a substantial and statistically significant predictive signal that is entirely absent from the standard simulation method.

\subsection{Limitations and Scalability to Genome-Scale Models (GEMs)}
The primary limitation of this study is its application to the small, highly curated \texttt{e\_coli\_core} model. While this system provided a clean environment to test our core hypothesis, scaling this approach to a full GEM presents several anticipated challenges. The increased network complexity will inevitably increase the topological "noise," and the sheer number of genes creates a "curse of dimensionality," a well-known challenge in machine learning \citep{Bellman1961}. Therefore, while our findings are decisive for the core network, their direct translation to genome-scale performance requires further investigation.

\subsection{Future Directions: Towards an Integrative Model}
The limitations of our current model point toward a clear future direction: the creation of a hybrid, multi-omics predictive model. The logical next step is to augment the successful topological features from this study with orthogonal layers of biological data. Integrating data such as gene expression levels from transcriptomics, protein-protein interaction data from sources like the STRING database \citep{Szklarczyk2019}, or even FBA-derived flux values as features could provide the necessary context to improve the signal-to-noise ratio.
\section{Conclusion}
In this work, we have demonstrated that a machine learning model trained exclusively on the topological features of a metabolic network decisively outperforms the standard Flux Balance Analysis method for predicting gene essentiality in the E. coli core metabolism. Our results validate the central hypothesis that network architecture is a primary determinant of biological function. The superiority of our "structure-first" framework suggests that it can be a more robust predictor than functional simulation, especially in biological systems characterized by high levels of redundancy. This study establishes a new benchmark for in-silico essentiality prediction and provides a strong conceptual and methodological foundation for developing the next generation of predictive tools in systems biology and rational drug target design.

\bibliography{references}

\end{document}